\documentclass[useAMS,usenatbib]{mn2e}
\usepackage{times}
\usepackage{graphicx}
\usepackage{subfigure}
\usepackage{psfrag}
\usepackage{array,amsmath,amssymb}
\usepackage[usenames]{color}
\usepackage{mathrsfs}
\usepackage{tikz}

\newcommand{\nc}{\newcommand}

\newcommand{\be}{\begin{equation}}
\newcommand{\ee}{\end{equation}}
\newcommand{\bea}{\begin{eqnarray}}
\newcommand{\eea}{\end{eqnarray}}

\renewcommand\[{\left[}

\newcommand\lsim{\mathrel{\rlap{\lower4pt\hbox{\hskip1pt$\sim$}}
    \raise1pt\hbox{$<$}}}
\newcommand\gsim{\mathrel{\rlap{\lower4pt\hbox{\hskip1pt$\sim$}}
    \raise1pt\hbox{$>$}}}
\def\ee{\end{equation}}
\def\be{\begin{equation}}

\newcommand{\salt}[1]{{\sc SALT-II}}
\newcommand{\saltorig}[1]{{\sc SALT}}
\newcommand{\sifto}[1]{{\sc SiFTO}}
\newcommand{\mlcs}[1]{{\sc MLCS}}

\nc{\Cm}{\hat{C}}
\nc{\sigint}{\sigma_{\mu}^\text{int}}
\nc{\muth}{\mu^{\text{th}}}
\nc{\cnot}{c_\star}
\nc{\xnot}{x_\star}
\nc{\onesn}{\mathbf{1}_n}
\nc{\ul}[1]{\underline{#1}}
\nc{\diff}{{\mathcal{T}}}
\nc{\lcdm}[1]{$\Lambda$CDM}

\def\reff@jnl#1{{\rm#1\/}}
\def\aj{\reff@jnl{AJ}}                 
\def\araa{\reff@jnl{ARA\&A}}           
\def\apj{\reff@jnl{ApJ}}               
\def\apjl{\reff@jnl{ApJ}}              
\def\apjs{\reff@jnl{ApJS}}             
\def\ao{\reff@jnl{Appl.Optics}}        
\def\apss{\reff@jnl{Ap\&SS}}           
\def\aap{\reff@jnl{A\&A}}              
\def\aapr{\reff@jnl{A\&A~Rev.}}        
\def\aaps{\reff@jnl{A\&AS}}            
\def\azh{\reff@jnl{AZh}}               
\def\baas{\reff@jnl{BAAS}}             
\def\jcap{\reff@jnl{JCAP}}           
\def\jrasc{\reff@jnl{JRASC}}           
\def\memras{\reff@jnl{MmRAS}}          
\def\mnras{\reff@jnl{MNRAS}}           
\def\pra{\reff@jnl{Phys.Rev.A}}        
\def\prb{\reff@jnl{Phys.Rev.B}}        
\def\prc{\reff@jnl{Phys.Rev.C}}        
\def\prd{\reff@jnl{Phys.Rev.D}}        
\def\prl{\reff@jnl{Phys.Rev.Lett}}     
\def\pasp{\reff@jnl{PASP}}             
\def\pasj{\reff@jnl{PASJ}}             
\def\qjras{\reff@jnl{QJRAS}}           
\def\skytel{\reff@jnl{S\&T}}           
\def\solphys{\reff@jnl{Solar~Phys.}}   
\def\sovast{\reff@jnl{Soviet~Ast.}}    
\def\ssr{\reff@jnl{Space~Sci.Rev.}}    
\def\zap{\reff@jnl{ZAp}}               
\def\nat{\reff@jnl{Nature}}            

\title[Constraining dark matter haloes using supernovae]{Bayesian constraints 
on dark matter halo properties using gravitationally-lensed supernovae} 
\author[N.V.~Karpenka et al.] 
{N.V.~ Karpenka$^1$\thanks{E-mail: nkarp@fysik.su.se}, M.C.~March$^2$, 
F.~Feroz$^3$ 
and M.P.~Hobson$^3$\\ 
$^1$The Oskar Klein Centre for Cosmoparticle Physics, Department of Physics, Stockholm University, AlbaNova, SE-106 91 Stockholm, Sweden\\
$^{2}$Astronomy Centre, University of Sussex, Brighton BN1 9QH, UK\\
$^3$Astrophysics Group, Cavendish Laboratory, JJ Thomson Avenue, Cambridge CB3 0HE, UK}

\date{Accepted 2012 December 21.  Received 2012 December 20; in original form 2012 July 16}
\pagerange{\pageref{firstpage}--\pageref{lastpage}}
\pubyear{2012}

\begin{document}
\label{firstpage}
\maketitle
\begin{abstract}
A hierarchical Bayesian method is applied to the analysis of Type-Ia
supernovae (SNIa) observations to constrain the properties of the dark
matter haloes of galaxies along the SNIa lines-of-sight via their
gravitational lensing effect. The full joint posterior distribution of
the dark matter halo parameters is explored using the nested sampling
algorithm {\sc MultiNest}, which also efficiently calculates the
Bayesian evidence, thereby facilitating robust model comparison.  We
first demonstrate the capabilities of the method by applying it to
realistic simulated SNIa data, based on the real 3-year data release
from the Supernova Legacy Survey (SNLS3). Assuming typical values for
the parameters in a truncated singular isothermal sphere (SIS) halo
model, we find that a catalogue analogous to the existing SNLS3 data
set is typically incapable of detecting the lensing signal, but a
catalogue containing approximately three times as many SNIa can
produce robust and accurate parameter constraints and lead to a clear
preference for the SIS halo model over a model that assumes no
lensing.  In the analysis of the real SNLS3 data, contrary to previous
studies, we obtain only a very marginal detection of a lensing signal
and weak constraints on the halo parameters for the truncated SIS
model, although these constraints are tighter than those typically
obtained from equivalent simulated SNIa data sets. This difference is
driven by a preferred value of $\eta \approx 1$ in the assumed
scaling-law $\sigma \propto L^\eta$ between velocity dispersion and
luminosity, which is somewhat higher than the canonical values of
$\eta = \tfrac{1}{4}$ and $\eta = \tfrac{1}{3}$ for early and
late-type galaxies, respectively.
\end{abstract}


\section{Introduction}\label{sec:intro}

In using Type Ia supernovae (SNIa) as `standardizable' candles to
constrain cosmological parameters, one typically assumes that the universe
is homogeneous and isotropic, and therefore one ignores gravitational
lensing effects due to cosmic structure along the line-of-sight to
each SNIa. In conventional cosmological SNIa analyses, this effect is
usually regarded as an additional source of uncertainty, which adds
extra scatter to the brightness of SNIa that increases with redshift
(\citealt{kantow95}, \citealt{freiman}, \citealt{wamb97},
\citealt{holzwald}, \citealt{lars00}). Fortunately, owing to flux
conservation, the effects of gravitational magnification and
demagnification average out and are therefore expected to lead to
negligible bias in cosmological parameter estimates (see, e.g.,
\citealt{Sarkar08}, \citealt{jacob08}).

Nonetheless, the gravitational lensing of SNIa can itself be used to
constrain cosmology (\citealt{metcalf99}, \citealt{dodelson06},
\citealt{zentner09}) and/or the properties of the lensing matter
(\citealt{rauch91}, \citealt{metcSilk}).  In the latter case, one
performs a complementary analysis to cosmological parameter estimation
by instead assuming a particular background cosmological model
(i.e. fixing the cosmological parameters to some concordance values) and
using the observed distance moduli to constrain the nature of the
cosmic structure, such as the properties of dark matter haloes, along
the lines-of-sight to the SNIa.  In principle, one might even hope to
perform a joint analysis to constrain the background cosmological
parameters and the nature of the cosmic structure simultaneously, but
such an approach is likely to suffer from strong degeneracies between
parameters.

An early tentative detection of gravitational lensing of SNIa was made
by \cite{jacob07} using a sample from the Great Observatories Origins
Deep Survey (GOODS; \citealt{riess04}, \citealt{strol04},
\citealt{riess07}). More recently, \cite{kronborg10} focussed on the
detection of a gravitational lensing signal by assuming the properties
of the dark matter haloes, fixing all the parameter values in the halo
model to `reasonable' values, and reported a positive result at the 99
per cent confidence level. Moreover, \cite{jonssonGOODS} used 24
high-redshift ($0.4 \la z \la 1.8$) SNIa from GOODS to constrain the
properties of dark matter haloes of galaxies also contained within
GOODS. This study was extended in \cite{jonssonSNLS} by using 175
high-redshift ($0.1 \la z \la 1$) SNIa from the 3-year data release of
the Supernova Legacy Survey (SNLS3; \citealt{astier06}) to constrain
the haloes of galaxies in the deep Canada-France-Hawaii Telescope
Legacy Survey (CFHTLS) fields. Although the SNIa in SNLS are typically
not as distant as those from GOODS, they are far more numerous and
selected in a more homogeneous way. \cite{jonssonSNLS} report the
detection of a gravitational lensing signal at the 92 per cent
confidence level, and place weak constraints on the parameters in their
halo model.

In this paper, we also use high-redshift SNIa from SNLS3 to constrain
the properties of dark matter haloes of galaxies in the CFHTLS fields
that intersect the SNIa lines-of-sight.  Our statistical methodology
differs greatly, however, from that used by
\citet{jonssonGOODS,jonssonSNLS} and other previous studies. As
recently discussed by \citet{march11}, the usual $\chi^2$-method used
to constrain cosmological parameters and/or the nature of cosmic
structure from lightcurve fits to SNIa observations (see
e.g. \citealt{astier06}; \citealt{kowalski}; \citealt{conley11})
suffers from some shortcomings in terms of its statistical
foundations and robustness, including not allowing for rigorous model
checking and not providing a reliable framework for the evaluation of
systematic uncertainties. Consequently, we instead analyse the \salt{}
lightcurve fits of the SNIa observations using the
statistically-principled and rigorous Bayesian hierarchical method
(BHM) of \citet{march11} to obtain a robust effective likelihood
function giving the probability of obtaining the observed SNIa data
(i.e. the parameter values obtained in the \salt{}
lightcurve fits) as a function of the parameters of the dark matter
halo model assumed for the galaxies along the lines-of-sight to the
SNIa. Moreover, rather than exploring the parameter space of the dark
matter halo model using simple gridding methods (see
e.g. \citealt{jonssonGOODS,jonssonSNLS}), we instead sample from the
full joint posterior distribution of the dark matter halo parameters
using a nested sampling algorithm
\citep{skilling04,feroz08,multinest}. This enables us to explore all
the halo model parameters simultaneously and allows for
straightforward marginalisation over subsets of them. The algorithm
also efficiently calculates the Bayesian evidence, thereby
facilitating robust model comparison.

The outline of this paper is as follows. In Section~2 we give a brief
summary of Bayesian inference methods, followed in Section~3 by a
description of our Bayesian methodology for using gravitational
lensing of SNIa to constrain the properties of dark matter haloes of
the galaxies intersecting the lines-of-sight. In Section~4, we
describe the SNLS3 supernovae and galaxies data used in our analysis.
We test the performance of our Bayesian methodology in Section~5 by
applying it to realistic simulated data based on the real SNLS3 data,
before analysing the real data sets and presenting our results in
Section~6. We give our conclusions in Section~7.

Throughout the paper, we assume a spatially-flat concordance
$\Lambda$CDM background cosmology, characterised by the parameters
${\cal C}\equiv\{\Omega_{\rm
  m,0},\Omega_{\Lambda,0},H_0\}=\{0.27,0.73,0.7\}$. Finally, we note
that this paper may be considered as complementary to our companion
paper \citep{Paper1}, in which we use the BHM to analyse the SNLS3
catalogue, together with additional SNIa, particularly at low
redshift, to constrain the background cosmological model, assuming no
gravitational lensing along the lines-of-sight to the SNIa.
 
\section{Bayesian inference}\label{sec:meth:bayes}

Our analysis methodology is built upon the principles of Bayesian
inference, which provide a consistent approach to the estimation of a
set of parameters $\mathbf{\Theta}$ in a model (or hypothesis) $H$
for the data $\mathbf{D}$. Bayes' theorem states that
\begin{equation} 
\Pr(\mathbf{\Theta}|\mathbf{D}, H) = \frac{\Pr(\mathbf{D}|\,\mathbf{\Theta},H)\Pr(\mathbf{\Theta}|H)}{\Pr(\mathbf{D}|H)},
\end{equation}
where, for brevity, we denote $\Pr(\mathbf{\Theta}|\mathbf{D}, H)
\equiv P(\mathbf{\Theta})$ as the posterior probability distribution
of the parameters, $\Pr(\mathbf{D}|\mathbf{\Theta}, H) \equiv
\mathcal{L}(\mathbf{\Theta})$ as the likelihood,
$\Pr(\mathbf{\Theta}|H) \equiv \pi(\mathbf{\Theta})$ as the prior, and
$\Pr(\mathbf{D}|H) \equiv \mathcal{Z}$ as the Bayesian evidence.

In parameter estimation, the normalising evidence factor is usually
ignored, since it is independent of the parameters $\mathbf{\Theta}$,
and inferences are often obtained by taking samples from the (unnormalised)
posterior using standard MCMC sampling methods, where at equilibrium
the chain contains a set of samples from the parameter space
distributed according to the posterior. This posterior constitutes the
complete Bayesian inference of the parameter values, and can be
marginalised over each parameter to obtain individual parameter
constraints.

In contrast to parameter estimation problems, for model selection the
evidence takes the central role and is simply the factor required to
normalize the posterior over $\mathbf{\Theta}$,
\begin{equation}
\mathcal{Z} = \int{\mathcal{L}(\mathbf{\Theta})\pi(\mathbf{\Theta})}\mathrm{d}^D\mathbf{\Theta},
\label{eq:3}
\end{equation} 
where $D$ is the dimensionality of the parameter space. As the average
of the likelihood over the prior, the evidence is larger for a model
if more of its parameter space is likely and smaller for a model with
large areas in its parameter space having low likelihood values, even
if the likelihood function is very highly peaked. Thus, the evidence
automatically implements Occam's razor. The question of model
selection between two models $H_{0}$ and $H_{1}$ can then be decided
by comparing their respective posterior probabilities given the
observed data set $\mathbf{D}$, as follows
\begin{equation}
R = \frac{\Pr(H_{1}|\mathbf{D})}{\Pr(H_{0}|\mathbf{D})}
  = \frac{\Pr(\mathbf{D}|H_{1})\Pr(H_{1})}{\Pr(\mathbf{D}| H_{0})\Pr(H_{0})}
  = \frac{\mathcal{Z}_1}{\mathcal{Z}_0} \frac{\Pr(H_{1})}{\Pr(H_{0})},
\label{eq:3.1}
\end{equation}
where $\Pr(H_{1})/\Pr(H_{0})$ is the a priori probability ratio for
the two models, which can often be set to unity but occasionally
requires further consideration.

Evaluation of the multidimensional integral in Eq. (\ref{eq:3}) is a
challenging numerical task. Standard techniques like thermodynamic
integration are extremely computationally intensive which makes
evidence evaluation at least an order of magnitude more costly than
parameter estimation. Some fast approximate methods have been used for
evidence evaluation, such as treating the posterior as a multivariate
Gaussian centred at its peak (see e.g. \citealt{hobson03}), but this
approximation is clearly a poor one for multimodal posteriors (except
perhaps if one performs a separate Gaussian approximation at each
mode). The Savage-Dickey density ratio has also been proposed (see
e.g. \citealt{trotta05}) as an exact, and potentially faster, means of
evaluating evidences, but is restricted to the special case of nested
hypotheses and a separable prior on the model parameters. Various
alternative information criteria for astrophysical model selection are
discussed by \citet{liddle07}, but the evidence remains the preferred
method.

The nested sampling approach, introduced by \citet{skilling04}, is a
Monte Carlo method targeted at the efficient calculation of the
evidence, but also produces posterior inferences as a
by-product. \citet{feroz08} and \citet{multinest} built on this nested
sampling framework and have introduced the {\sc MultiNest} algorithm
which is very efficient in sampling from posteriors that may contain
multiple modes and/or large (curving) degeneracies and also calculates
the evidence. This technique greatly reduces the computational cost 
of Bayesian parameter estimation and model selection and has already 
been applied to a number of problems in astrophysics (see 
e.g. \citealt{2011MNRAS.tmp.2013F, 2011MNRAS.415.3462F, 
2009MNRAS.400.1075B, 2010CQGra..27g5010F}). We employ this technique in 
this paper. 

\section{Analysis methodology}\label{sec:meth}

\subsection{Definition of the SNIa data}\label{sec:meth:defSNIa}

In practice, there are no perfect astronomical standard candles.
In particular, SNIa have absolute magnitudes that
vary by about $\pm 0.8$~mag in the $B$-band due to physical
differences in how each supernova is triggered and also due to
absorption by its host galaxy. Nonetheless, SNIa do constitute a set
of `standardizable' candles, since by applying small corrections to
their absolute magnitudes, derived by fitting multi-wavelengths
observations of their lightcurves, one can reduce the scatter
considerably, to around $\pm 0.15$~mag in the $B$-band.  In essence,
SNIa with broader light curves and slower decline rates are
intrinsically brighter than those with narrower light curves and fast
decline rates (\citealt{phillips93}). 

Several methods are available for fitting SNIa lightcurves (and
constraining cosmological parameters), including, amongst others, the
Multi-Colour Lightcurve Shape (MCLS) strategy \citep{JhaRiess2007},
CMAGIC \citep{wang03,conley06}, and the Spectrally Adaptive Lightcurve
Template (SALT) method \citep{GuyAstier2007}, the current version of
which is \salt{}. The relative merits of these methods is a topic of
much debate, but the \salt{} method is particularly attractive for our
purposes, since (unlike MCLS) it first fits each SNIa lightcurve to
obtain three parameters controlling the SN magnitude, stretch and
colour corrections to a template `learned' from nearby and distant SN;
only in a second, separate step are these fits used to constrain
cosmological parameters. Following \citet{march11}, we may therefore
use the products of the first step as the inputs to a statistically
rigorous Bayesian hierarchical model.

Our analysis takes place after the selection cuts, lightcurve fitting
and Malmquist correction have been performed.  For each selected SNIa,
in addition to an estimate $\hat{z}$ of its redshift and an associated
uncertainty $\sigma_z$, derived from observations of its host galaxy,
we take as our basic data the output from the \salt{} lightcurve
fitting algorithm, which produces the best-fit values:
$\hat{m}_{B}^\ast$, the rest frame $B$-band apparent magnitude of the
supernovae at maximum luminosity; $\hat{x}_1$, the stretch parameter
related to the width of the fitted light curve; and $\hat{c}$, the
colour excess in the $B$-band at maximum luminosity. These are
supplemented by the covariance matrix of the uncertainties in the
estimated lightcurve parameters, namely
\begin{equation}
\widehat{C} = 
\left(
\begin{array}{ccc}
\sigma^2_{m^\ast_B} & \sigma_{m^\ast_B,x_1} & \sigma_{m^\ast_B,c} \\[2mm]
\sigma_{m^\ast_B,x_1} & \sigma^2_{x_1} & \sigma_{x_1,c}\\[2mm]
\sigma_{m^\ast_B,c} & \sigma_{x_1,c} &  \sigma^2_{c} \\
\end{array}
\right).
\label{eq:covmat}
\end{equation}
Therefore, our basic input data for each SN
($i=1,\ldots,N_{\rm SN}$) are
\begin{equation}
D_i \equiv
\{\hat{z}_i,\hat{m}_{B,i}^\ast,\hat{x}_{1,i},\hat{c}_i\},
\label{eqn:inputdata}
\end{equation}
and we assume (as is implicitly the case throughout the SNe literature)
that the vector of values
$(\hat{m}_{B,i}^\ast,\hat{x}_{1,i},\hat{c}_i)$ for each SN is
distributed as a multivariate Gaussian about the true values, with
covariance matrix $\widehat{C}_i$. The `observed' distance modulus
$\mu^{\rm obs}_i$ for each SN is then modelled as
\begin{equation}
\mu_i^{\rm obs} = \hat{m}_{B,i}^\ast - M_i + \alpha \hat{x}_{1,i} - \beta\hat{c}_i,
\end{equation}
where $M_i$ is the (unknown) $B$-band absolute magnitude of the SN,
and $\alpha$, $\beta$ are (unknown) nuisance parameters (assumed the
same for all SN) controlling the stretch and colour corrections.

It should be noted that a sophisticated Bayesian hierarchical method
has recently been proposed by \citet{mandel09,mandel10} to fit
optical and infrared lightcurve data. This may provide a more robust
technique for defining the basic SNIa data that we use in our
subsequent analysis, but we leave the investigation of this issue to a
future work.

\subsection{Computing the predicted distance moduli}\label{sec:meth:magn}
In using SNIa to constrain cosmic structure, the predicted distance
modulus must include the effect of gravitational magnification due to
cosmic structure along the line-of-sight to each supernova. In
particular, we assume this magnification is due to dark matter haloes
associated with known galaxies intersecting the line-of-sight. Thus,
for each SNIa, the predicted magnification depends on the sets of
parameters $\{\bmath{g},\bmath{h}\}$, where
\begin{equation}
\bmath{g} = 
\{z_{\rm gal}^1,\btheta_{\rm gal}^1,M_{B}^1,\tau^1,\ldots,
z_{\rm gal}^{N_{\rm gal}},\btheta_{\rm gal}^{N_{\rm gal}},M_{B}^{N_{\rm gal}},\tau^{N_{\rm gal}}\}
\end{equation}
contains the redshift, sky position, absolute $B$-band magnitude and
spectral type of the $N_{\rm gal}$ galaxies that intersect the
line-of-sight to the SNIa, and $\bmath{h}$ contains the parameters of
the assumed dark matter halo model for these galaxies (see
Section~\ref{sec:meth:halo}). In general, the parameters $\bmath{g}$
are (naturally) different for each SN and are assumed known; one
wishes to place constraints on the unknown halo parameters
$\bmath{h}$, which (perhaps unrealistically) are assumed common to all
the foreground galaxies.

To compute the predicted magnification of a SNIa, we use the
weak-lensing approximation (see, e.g.. \citealt{schneider02}), the validity of which was checked by \citet{jonssonSNLS}
by comparing it with a ray-tracing algorithm.  For the SNIa sample
we consider here (see Section~\ref{sec:data:sn}), which have relatively low
redshifts, the weak-lensing approximation was found to be accurate to
within $\sim 5$ per cent. In this approximation, the predicted
distance modulus of a SNIa, expressed in terms of magnitudes, is
related to the convergence $\kappa$ along its line-of-sight by
\begin{equation}
\mu(z,{\cal C},\bmath{g},\bmath{h}) \approx \mu_0(z,{\cal C}) 
- 2.17[\kappa_{\rm los}(\bmath{g},\bmath{h})
-\kappa_{\rm b}(\bmath{h})],
\label{eq:mag-kappa}
\end{equation}
where $\mu_0(z,{\cal C})$ is the predicted distance modulus for our assumed
cosmological parameters ${\cal C}$, neglecting
gravitational lensing, $\kappa_{\rm los}(\bmath{g},\bmath{h})$ is a
sum over the contributions to the convergence from each galaxy along
the line-of-sight, so that $\kappa_{\rm los} = \sum_{j=1}^{N_{\rm
    gal}} \kappa^j_{\rm gal}$, and $\kappa_{\rm b}(\bmath{h})$
represents the compensating effect of the background density and acts
as a normalisation allowing the magnification relative to a
homogeneous universe to be computed (\citealt{jonssonSNLS}). Flux
conservation implies that $\langle \kappa \rangle = 0$ and to ensure
this condition is satisfied we set $\kappa_{\rm b}(\bmath{h}) =
\langle \kappa_{\rm los}(\bmath{g},\bmath{h}) \rangle$, where the
latter is calculated using a large number of randomly selected
lines-of-sight.
  
\subsection{Halo model}\label{sec:meth:halo}

Our primary goal is to constrain the parameters $\bmath{h}$ that
describe the properties of the dark matter haloes associated with the
galaxies along the lines-of-sight to the SNIa. We assume that each
galaxy contributes to the convergence by an amount
\begin{equation}
\kappa_{\rm gal}(\bxi)=\frac{\Sigma(\bxi)}{\Sigma_{\rm c}}, 
\label{eq:4}
\end{equation}
where the surface density, $\Sigma(\bxi)$, is obtained by projecting the matter distribution onto a lens plane,
\begin{equation}
\Sigma(\bxi)=\int_{-\infty}^{\infty} \rho(\bxi,y)\,{\rm d}y,
\label{eq:5}
\end{equation} 
where $\bxi$ is a vector in the plane and $y$ is a coordinate along
the line of sight; it is at this point that the density profile
$\rho(\bmath{r})$ of the dark matter halo enters the calculation.
The denominator in (\ref{eq:4}) is the critical
surface density,
\begin{equation}
\Sigma_{\rm c}=\frac{c^2}{4\pi G}\frac{D_{\rm s}}{D_{\rm l}D_{\rm ls}},
\label{eq:6}
\end{equation}
which, in turn, depends on the angular diameter distances between the
observer and the source, $D_{\rm s}$, the observer and the lens,
$D_{\rm l}$, and the lens and the source, $D_{\rm ls}$.  These distances
are computed from the redshifts of the SN and galaxy respectively,
assuming our concordance background cosmology.

In order to compare our results directly with those of
\cite{jonssonGOODS, jonssonSNLS}, in this paper we assume
that the density profile $\rho(\bmath{r})$ of the dark matter halo is
described by a truncated singular isothermal sphere (SIS), although we
note that our approach could be straightforwardly extended to consider
alternative halo models, such as the Navarro--Frenk--White (NFW)
profile (\citealt{navarro97}). The radial density distribution of a
singular isothermal sphere (SIS) is given by
\begin{equation}
\rho(r)=\frac{\sigma^2}{2\pi G}\frac{1}{r^2},
\end{equation}
which depends on the single free parameter $\sigma $, the
one-dimensional velocity dispersion of its constituent particles.
Since the SIS profile has a divergent total mass, we truncate it at a
radius $r=r_{\rm t}$, which is thus a second free parameter.

The surface density of a truncated SIS is easily calculated using
(\ref{eq:5}) and found to be
\begin{equation}
\Sigma(\xi)=\left\{ 
\begin{array}{ll}
\frac{\sigma^2}{\pi G\xi}\arctan\sqrt{ {r_{\rm t}^2}/{\xi^2}-1} &\mbox{ if $\xi \leq r_{\rm t}$} \\[2mm]
 0  &\mbox{ if $\xi > r_{\rm t}$},
       \end{array}
\right.
\label{eq:sissigma}
\end{equation}
which can, in turn, be substituted into (\ref{eq:4}) to obtain the
convergence $\kappa_{\rm gal}(\xi)$ due to the galaxy dark matter
halo.

To allow for and investigate the relationship between galaxy
luminosity and velocity dispersion, we follow \cite{jonssonSNLS} and
adopt the Faber--Jackson \citep{FG76} and Tully--Fisher-like \citep{TF77} 
scaling law
\begin{equation}
\sigma=\sigma_{*}\left( \frac{L}{L_*} \right)^{\eta},
\label{eq:sislaw}
\end{equation}
where $L_*$ is a fiducial luminosity, which we take to be
\mbox{$L_*=10^{10}h^{-2}L_{\sun}$} in the $B$-band, and $\sigma_\ast$
is the corresponding fiducial velocity dispersion. In terms of
absolute $B$-band magnitudes, with which we will be working, the
scaling relation becomes
\begin{equation}
\sigma=\sigma_*10^{-\eta(M_B-M_B^*)/2.5}, 
\end{equation}
where $M_B^*=-19.52+5\log_{10} h$ and $h$ is the Hubble
constant in units of 100 km s$^{-1}$ Mpc$^{-1}$. The (aperture) mass-to-light ratio of the galaxies is determined by $\eta$, since
\begin{equation}
\frac{M(r \leq R)}{L} = \frac{2\sigma_\ast^2 (L/L_\ast)^{2\eta}R}{GL}
\propto L^{2\eta-1},
\end{equation}
provided $R \leq r_t$. Thus, for example, one has a constant
mass-to-light ratio $M/L \propto L^0$ if $\eta=0.5$.

For the truncation radius, we again follow \cite{jonssonSNLS} and
assume it obeys a scaling law of the form\footnote{One could 
assume a $r_{\rm t}$--$L$ relationship, to break the coupling
between $\eta$ and $\gamma$, but this would make comparison 
with \cite{jonssonSNLS} difficult.
}
\begin{equation}
r_{\rm t}=r_*\left(\frac{\sigma}{\sigma_*}\right)^{\gamma}=r_*\left(\frac{L}{L_*}\right)^{\eta\gamma},
\label{eq:trlaw}
\end{equation}
where $r_\ast$ is a fiducial truncation radius. Since we only include
galaxies located a distance $\theta_{\rm c}$ from the position of the
SN~Ia, the truncation radius has an effect only if $r_{\rm t}/D_{\rm
  d}<\theta_{\rm c}$. For $\theta_{\rm c}=60\arcsec$ and $z_{\rm
  gal}<1$ this corresponds to $r_{\rm t} \la 300h^{-1}$
kpc.

Thus, for the truncated SIS halo model, we wish to constrain the four
parameters $\bmath{h}=\{\gamma,\eta,\sigma_\ast,r_\ast\}$.

\subsection{Likelihood function}\label{sec:meth:like}

To construct the likelihood function for the SNIa data, we adopt the
Bayesian hierarchical model proposed by \citet{march11}. This has been
shown to deliver tighter constraints on cosmological parameters than
the usual $\chi^2$-method, and simultaneously provides a robust
statistical framework for the full propagation of systematic
uncertainties to the final inferences.

We require the likelihood of the input data (\ref{eqn:inputdata}) for
our full catalogue of SNe, given the parameters of our model, namely
\begin{equation}
\Pr(\hat{\bmath{m}}_{B}^\ast,\hat{\bmath{x}}_{1},\hat{\bmath{c}},\hat{\bmath{z}}|\bmath{h},\alpha,\beta,
\sigma_{\rm int}),
\label{eqn:likelihood}
\end{equation}
which also depends on the cosmological parameters ${\cal C}$, the
covariance matrices $\widehat{C}_i$ 
of the uncertainties on the input
data $(\hat{m}_{B,i}^\ast,\hat{x}_{1,i},\hat{c}_i)$ for each SN, and the
uncertainties $\sigma_{z,i}$ in the estimated redshifts $\hat{z}_i$, all
of which are assumed known. In particular, we seek to constrain the
unknown halo parameters $\bmath{h}$, global colour and stretch correction
multipliers $\alpha$ and $\beta$, and the intrinsic dispersion
$\sigma_{\rm int}$ of SNIa absolute magnitudes (all of these are
assumed to be `global' parameters, i.e. common to every SN). 

Following \citet{march11}, we compute the likelihood
(\ref{eqn:likelihood}) by first introducing for each SN the hidden
variables $M_i$, $x_i$, $c_i$ and $z_i$, which are, respectively, the
true (unknown) values of its absolute $B$-band magnitude, stretch and
colour corrections, and redshift. These are then assigned priors,
which themselves contain further nuisance parameters, and all the
introduced parameters are marginalised over to obtain the likelihood
(\ref{eqn:likelihood}). The details of this procedure are given in
Appendix B of our companion paper \citep{Paper1}.  By assuming
separable Gaussian priors on the hidden variables and nuisance
parameters, one can perform all the marginalisations analytically,
except for two nuisance parameters $R_x$ and $R_c$ (the dispersions of
the priors on the stretch and colour corrections, respectively) that
must be marginalised over numerically. The full likelihood function
thus depends on the parameters $\mathbf{\Theta} =
\{\bmath{h},\alpha,\beta,\sigma_{\rm int},R_x,R_c\}$, and is
9-dimensional for the truncated SIS halo model.

\subsection{Priors on the sampled parameters}\label{sec:meth:prior}

To determine the Bayesian inference problem completely, it only remains
to specify the prior $\pi(\mathbf{\Theta})$ on the parameters to be
sampled. The choice of prior is particularly important for weak
lensing analyses, since the problem is inherently underconstrained and
therefore any prior information available is extremely useful. One
should, however, be careful in the choice of priors not to impose too
strong assumptions, which may lead to erroneous inferences.

Following \citet{march11}, we adopt the separable priors listed in
Table~\ref{tab:priors1} on each of the `non-halo' (nuisance)
parameters $\sigma_{\rm int}$, $\alpha$, $\beta$, $R_x$ and $R_c$, and
on the halo parameters $\bmath{h}$ for the truncated SIS model; these
correspond to broad, conservative assumptions.
\begin{table}
\begin{center}
\begin{tabular}{lcr}
\hline
Parameter & Symbol & Prior \\
\hline
{\em Nuisance parameters} & & \\
Dispersion of absolute magnitude & $\sigma_{\rm int}$ &  
${\cal U}(-3,0)$ on $\ln\sigma_{\rm int}$\\
Stretch multiplier & $\alpha$ & ${\cal U}(0,4)$ \\
Colour multiplier & $\beta$ & ${\cal U}(0,4)$ \\
Dispersion of stretch correction & $R_x$ &  ${\cal U}(-5,2)$ on $\ln R_x$ \\
Dispersion of colour correction & $R_c$ &  ${\cal U}(-5,2)$ on $\ln R_c$\\[1mm]
{\em Halo parameters} & & \\
Fiducial velocity dispersion (km s$^{-1}$) \hspace*{-0.4cm} & $\sigma_\ast$ & ${\cal U}(0,300)$ \\
Fiducial truncation radius (Mpc) & $r_\ast$ &  ${\cal U}(0,0.4)$\\
Exponent of $\sigma$--$L$ power law & $\eta$ & ${\cal U}(-2,2)$ \\
Exponent of $r_t$--$\sigma$ power law & $\gamma$ &  ${\cal U}(-5,5)$ \\
\hline
\end{tabular}
\caption{Priors on the nuisance parameters and the halo parameters of the truncated SIS model, where ${\cal U}(a,b)$ denotes
  a uniform distribution between the limits $a$ and $b$.\label{tab:priors1}}
\end{center}
\end{table}
%

\section{Supernovae and galaxies data sets}\label{sec:data}

To apply the Bayesian analysis methodology described above to
constrain the dark matter haloes of galaxies, we need observations
both of supernovae and foreground galaxies. As in J\"onsson et
al. (2010b), we use high-redshift ($0.1 \la z \la 1$) SNIa from the
3-year SNLS data set \citep{guy09} to constrain the properties of dark
matter haloes of galaxies in the CFHTLS fields that intersect the SNIa
lines-of-sight.

The SNLS consists both of photometric and spectroscopic observations.
The photometry is obtained as part of the deep component of CFHTLS
with the one square-degree imager MegaCam (\citealt{boul03}). The deep
part of CFHTLS comprises four fields (D1, D2, D3 and D4), each
$\approx$ 1 deg$^2$ in size, imaged in $u^\ast$, $g'$, $r'$, $i'$ and
$z'$ filters approximately every 4--5 days during dark and grey time,
suitable for detecting supernovae and building light curves
(\citealt{astier06}; \citealt{guy09}). The spectroscopic observations
are used to determine the nature of the supernovae candidates and
measure their redshifts (\citealt{howell05}; \citealt{bron08};
\citealt{ball09}).

\subsection{Supernovae}\label{sec:data:sn}

Our initial sample consisted of 230 high-redshift ($0.1 \la z \la 1$)
SNIa from the 3-year SNLS data set \citep{guy09}. We note that this is
13 fewer than available to \cite{jonssonSNLS}, since these SNIa were
excluded from the \salt{} fits for various reasons described in
\cite{conley11}, and do not appear in the tables in that paper. The
sample of 230 SNIa is further reduced, however, by edge effects and by
some parts of the deep CFHTLS fields being covered by bright stars
that have to be masked \citep{sul06}. SNIa located too close to the
boundary of the field or to a masked region are removed from the
sample because of the lack of observations of foreground
galaxies. Details of this procedure are given in \cite{jonssonSNLS}.
Only 162 of the initial sample of 230 SNIa fulfilled this selection
criterion.

To compute the predicted magnification due to the foreground galaxies
along the line-of-sight to each SNIa, we require an estimate of its
sky location $\btheta_{\rm SN}$ and redshift $\hat{z}_{\rm SN}$. The
sky location is obtained from the CFHTLS $i'$-band photometry. When
available, $\hat{z}$ is taken to be the spectroscopic redshift of the
SNIa host galaxy; otherwise the spectroscopic redshift of the SNIa
itself is used.

Selection cuts, \salt{} lightcurve fitting and Malmquist corrections
are made by the SNLS3 team and are already implemented in the supplied
data files.  As described in Section~\ref{sec:meth:defSNIa}, the
\salt{} fitting algorithm is applied to each SNIa lightcurve to obtain
best-fit estimates of: $\hat{m}_{B}^\ast$, the rest frame $B$-band
apparent magnitude of the supernovae at maximum luminosity;
$\hat{x}_1$, the stretch parameter of the fitted light curve; and
$\hat{c}$, the colour excess in the $B$-band at maximum
luminosity. These are supplemented by the covariance matrices
$\widehat{C}$ of the uncertainties in the estimated lightcurve
parameters, which are taken from \cite{conley11}.

\subsection{Galaxies}\label{sec:data:gal}

Our foreground galaxies are taken from the SNLS galaxy catalogues in
the deep CFHTLS fields. As discussed in Section~\ref{sec:meth:magn},
to perform our analysis we require for each galaxy: the redshift
$z_{\rm gal}$, sky position $\btheta_{\rm gal}$, absolute $B$-band
magnitude $M_B$, and spectral type $\tau$.  The techniques used to
obtain these galaxy properties are described in detail in
\citet{sul06} and \citet{jonssonSNLS}.

\section{Application to simulated supernovae data}\label{sec:sim}

In order to test our Bayesian analysis methodology, we first apply it
to simulated SNIa data.  First, SNIa photometric data in the absence
of gravitational lensing were simulated and fitted using the publicly
available SNANA package \citep{Kessler2009SNANA}, in a identical
manner to that described in our companion paper \citep{Paper1}. In
summary, the data were simulated to match closely the SNLS3 data set
\citep{GuySullivan2010} by using the SNLS3 co-added simulation library
files (which are publicly available as part of the SNANA package), a
coherent magnitude smearing of $0.12$, and colour smearing. The colour
smearing effect, or broad-band colour dispersion model, implemented in
the data simulation is the EXPPOL model described by fig.~8 of
\cite{GuySullivan2010}, and the simulated Malmquist bias is based on
fig.~14 of \cite{PerrettBalam2010}.

The SNANA SNIa data simulation is a two-stage process that mimics the
real data collection and analysis process. The first stage is the
simulation of photometric data in accordance with the characteristic
instrument and survey properties of the SNLS3 survey using the SNLS3
simulation library files mentioned above.  The second stage is the
lightcurve fitting process in which the photometric data are fitted to
\salt{} templates to give estimates of the SNIa absolute B-band
magnitude $\hat{m}_B$, lightcurve stretch $\hat{x}_1$ and colour
$\hat{c}$. At this lightcurve fitting stage, basic cuts are made to
discard SNIa with a low signal-to-noise ratio and/or too few observed
epochs in sufficient bands. After the lightcurve fitting stage we make
a redshift dependent magnitude correction to correct for the Malmquist
bias; the correction is taken from a spline interpolation of table 4
in \cite{PerrettBalam2010}. 

The resulting simulated SNIa are randomly distributed on the sky
across the four SNLS3 fields, outside of our masked regions, and have
a redshift distribution appropriate for the SNLS3 survey; in total we
simulate $10^4$ SNIa. The final stage of our simulation process adds
the lensing contribution due to the (assumed) galaxy haloes along the
line-of-sight to each SN. This is performed using the {\em real} SNLS
galaxy catalogue, by assuming each galaxy to have a halo described by
a truncated SIS profile with the parameter values listed in
Table~\ref{tab:sims}. The choice of the velocity dispersion
$\sigma_\ast$ and truncation radius $r_\ast$ correspond to the
best-fit values obtained by \cite{jonssonSNLS}.  The value
$\eta=\tfrac{1}{3}$ corresponds to the Tully--Fisher relation for
late-type galaxies, which should make up the majority of the SNLS
sample, since nearly 95 per cent of them are star-forming galaxies. We
also assume $\gamma=1$, which corresponds to the truncation radius
being linearly proportional to $\sigma$, and has the natural
consequence that the halo mass scales as $M \propto \sigma^3$. For an
assumed set of halo parameters, the lensing contribution from the
haloes along the line-of-sight to each simulated SN is then added to
the SN apparent magnitude, and its uncertainty is left unchanged.

\begin{table}
\begin{center}
\begin{tabular}{lcr}
\hline
Halo parameter & Symbol & Value \\
\hline
Fiducial velocity dispersion (km s$^{-1}$) & $\sigma_\ast$ & 120  \\
Fiducial truncation radius (Mpc) & $r_\ast$ & 0.07  \\
Exponent of $\sigma$--$L$ power law & $\eta$ & $\tfrac{1}{3}$ \\
Exponent of $r_t$--$\sigma$ power law & $\gamma$ & 1  \\[1mm]
\hline
\end{tabular}
\caption{Halo parameter values used in the generation of simulated
  gravitationally-lensed SNIa data.\label{tab:sims}}
\end{center}
\end{table}

\subsection{Characteristics of the lensing signal}
\label{sec:character}

Before we apply our BHM to sets of simulated supernovae data, it is of
interest first to investigate the general characteristics of the
simulated lensing signal. It is worth reiterating that this signal is
calculated for each of the $10^4$ simulated SNIa using the {\em true}
galaxies from the SNLS catalogue along each line-of-sight, albeit with
the assumption that each such galaxy has a dark matter halo described
by a truncated SIS profile with the fiducial parameter values given in
Table~\ref{tab:sims}.

\begin{figure}
\begin{center}
\includegraphics[width=0.8\columnwidth]{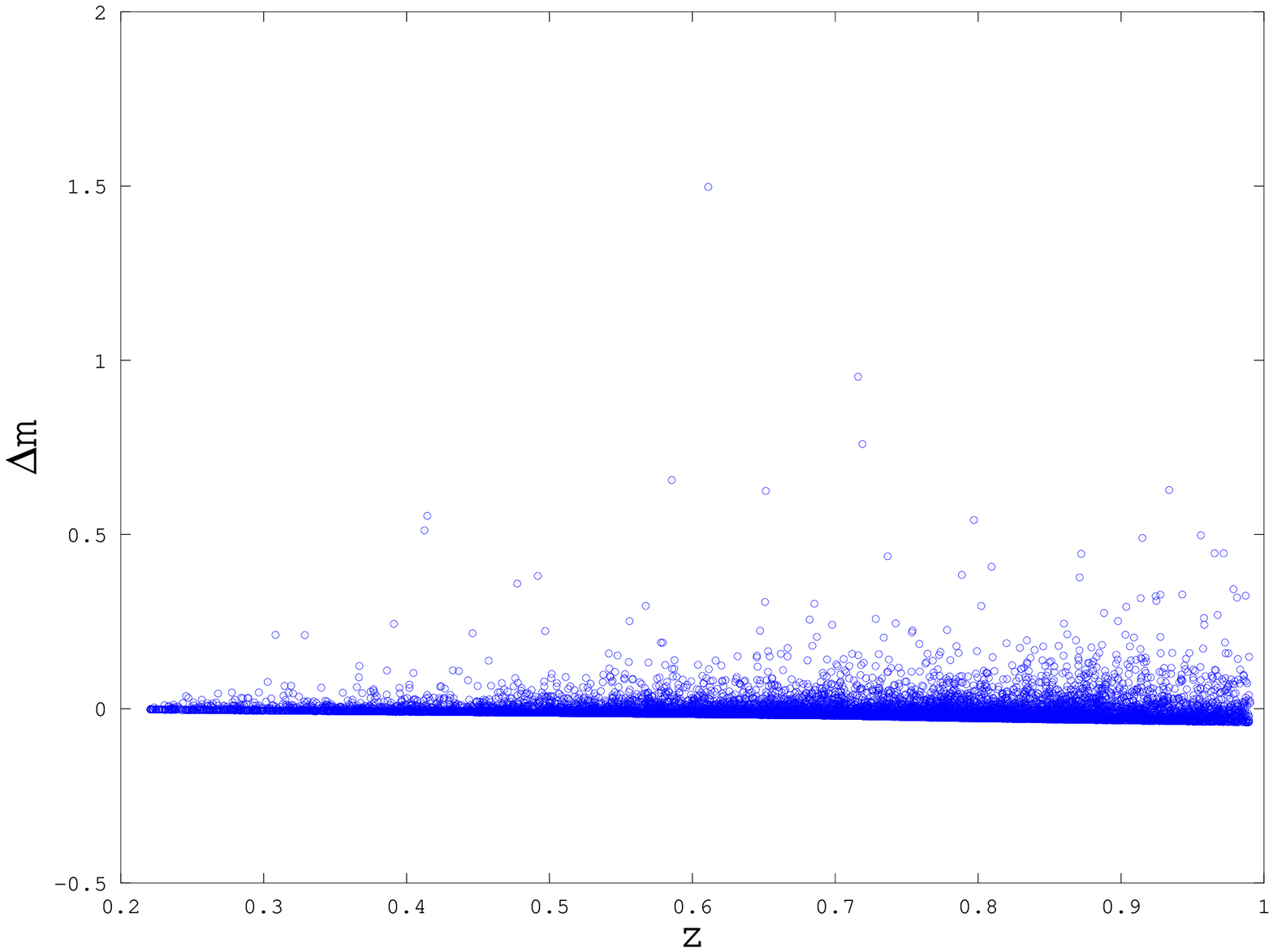}\\[1mm]
\includegraphics[width=0.8\columnwidth]{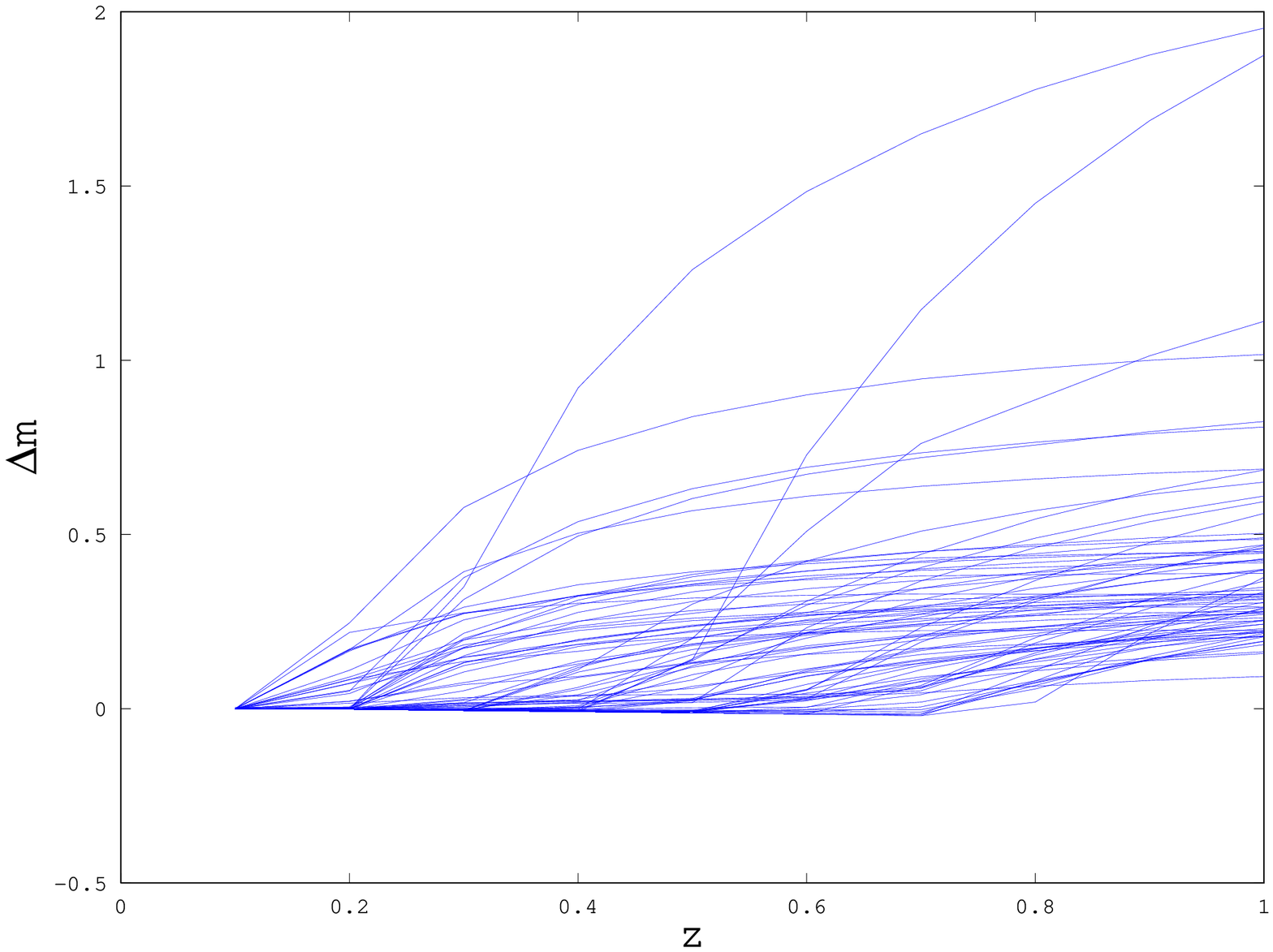}
\caption{Top: magnification of $10^4$ simulated SNIa randomly
  positioned in the true SNLS galaxy catalogue, and drawn from a
  redshift distribution appropriate for the SNLS3 supernovae survey. Bottom:
magnification versus redshift for the 75 lines-of-sight that
  exhibit the strongest simulated lensing effect.
  \label{fig:simmags}}
\end{center}
\end{figure}

In Fig.~\ref{fig:simmags} (top panel), we plot the magnification
factor for each of the $10^4$ simulated SNIa. We note, in particular,
that most of the SNIa are demagnified, as a result of the background
correction described in Section~\ref{sec:meth:magn}, but a small
number of SNIa are significantly magnified. When analysing random
samples of (say) a few hundred SNIa, one would therefore expect a wide
variation in the significance at which one detects a gravitational
lensing signal, depending on whether the sample contains one or more
of the SNIa that are strongly magnified.

We note further that, for the significantly lensed SNIa, there is no
clear correlation between the size of the magnification and redshift,
suggesting that high-redshift SNIa are not necessarily to be preferred
for detecting the lensing signal of dark matter haloes. It is of
interest to investigate further the lines-of-sight along which there
is a significant lensing effect. In Fig.~\ref{fig:simmags} (bottom
panel), we plot the magnification as a function of redshift along the
75 lines-of-sight that exhibit the highest magnification. Although the
magnification of the three most highly lensed lines-of-sight continues
to increase markedly up to $z=1$, the magnification along the
remaining lines-of-sight typically does not increase appreciably
beyond $z \sim 0.5$, again showing that high-redshift SNIa are not
necessarily of more use in detecting the lensing signal.

\subsection{Results for random samples of 162 supernovae}

\begin{figure}
\begin{center}
\includegraphics[width=0.95\columnwidth]{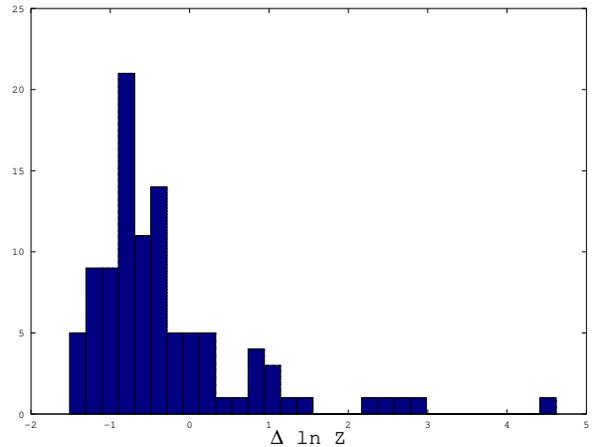}
\caption{Histogram of the log-evidence difference
  $\Delta\ln\mathcal{Z}$ between the SIS halo model and the null
  (no-lensing) model obtained from the analysis of one hundred
random samples of 162 SNIa.\label{fig:sis162hist}}
\end{center}
\end{figure}

In Fig.~\ref{fig:sis162hist}, we plot the histogram of the
log-evidence difference $\Delta\ln\mathcal{Z}$, for the SIS halo model
relative to the null (no-lensing) model, obtained from the analysis of
one hundred random samples of 162 SNIa (to match the number in the
real SNLS3 sample to be analysed in Section~\ref{sec:real}).
As anticipated, there is a large variation, with
$\Delta\ln\mathcal{Z}$ ranging from about $-1.5$ to $4.5$.  According
to Jeffreys' scale, the former corresponds to the SIS halo model being
`substantially' disfavoured, whereas as the latter indicates that it
is `very strongly' preferred, relative to the no-lensing model.  We
note that the histogram has a median $-0.6$, mean $-0.3$ and standard
deviation $1.0$.  We also find that the random samples having large
values of $\Delta\ln\mathcal{Z}$ typically contain a number of the
most strongly lensed SNIa.

\begin{figure*}
\begin{center}
\includegraphics[width=0.95\columnwidth]{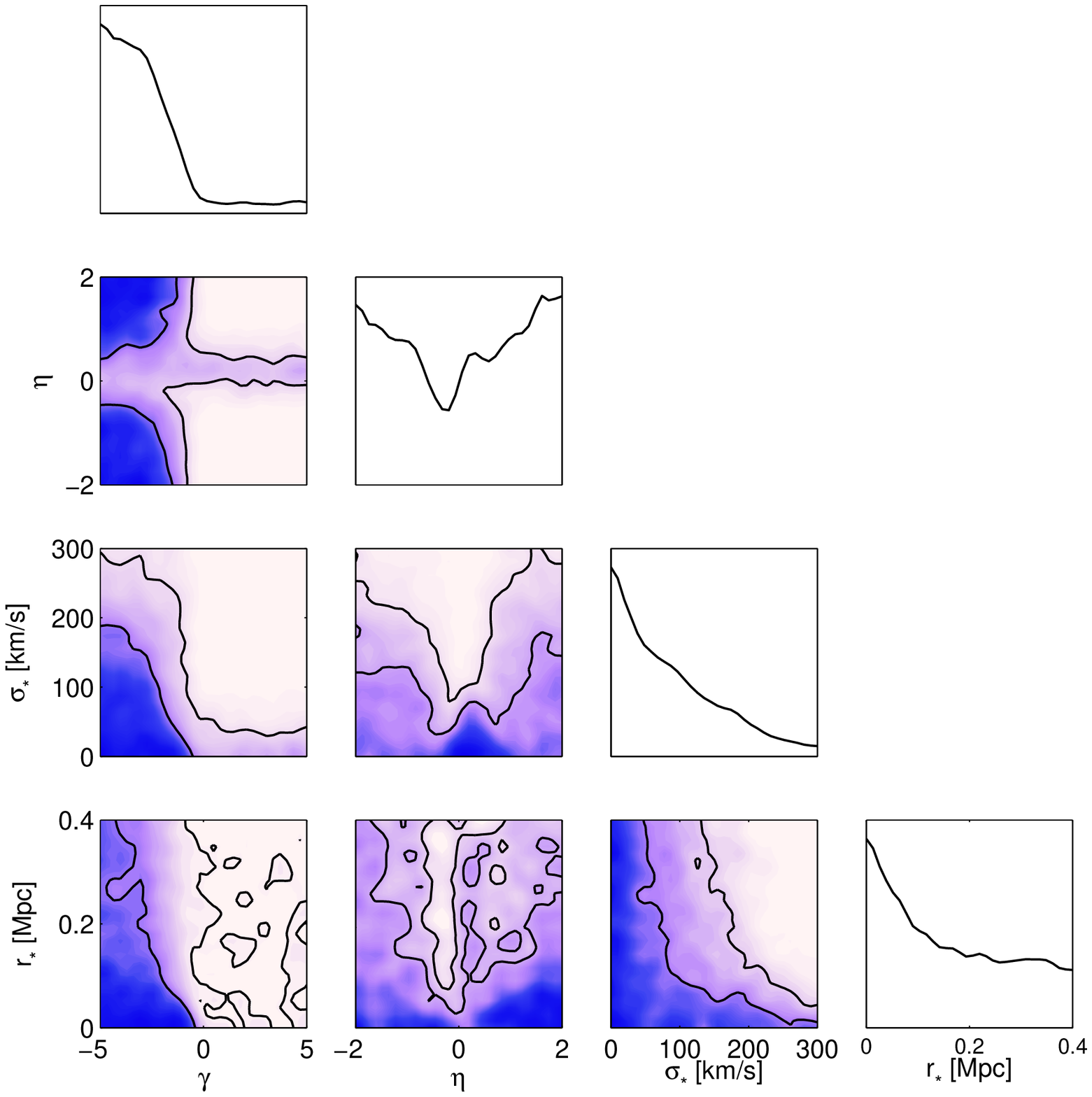}\qquad\qquad
\includegraphics[width=0.95\columnwidth]{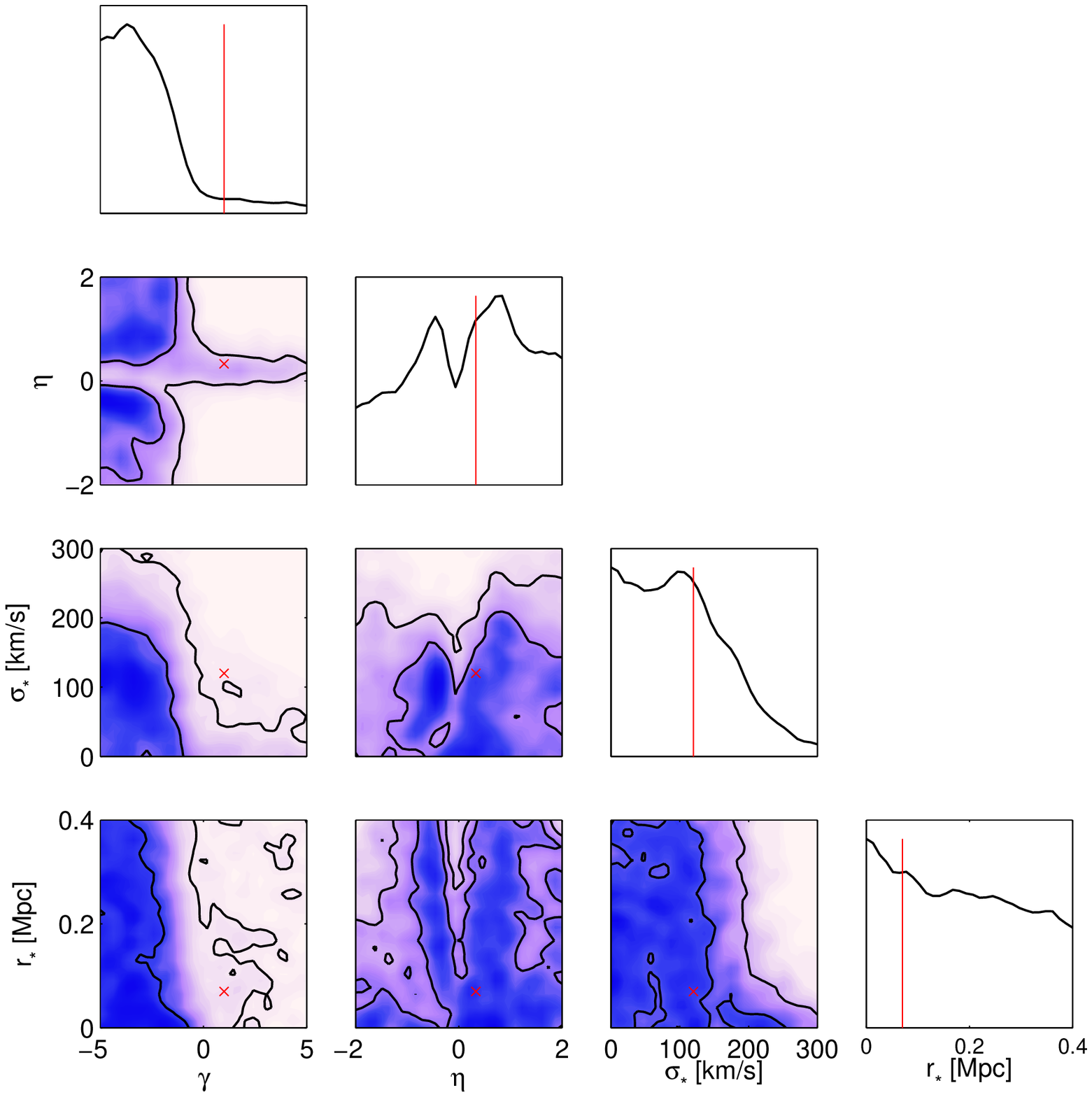}
\caption{1D and 2D marginalised posteriors distributions for the
  parameters $\bmath{h}=\{\gamma,\eta,\sigma_\ast,r_\ast\}$ of the
  truncated SIS halo model, derived from the analysis of 162 simulated
  SNIa data generated assuming no lensing (left) and a truncated SIS
  model (right). In the right-hand panel, true parameters are
  indicated by vertical lines and crosses in 1D and 2D plots
  respectively. The SNIa sample used corresponds to that with the
  median value $\Delta\ln\mathcal{Z}=-0.6$ from the histogram in
  Fig.~\ref{fig:sis162hist}. \label{fig:sim-tsis-1gal-162}}
\end{center}
\end{figure*}

For the random sample that yields the median value of
$\Delta\ln\mathcal{Z}=-0.6$, we plot the corresponding parameter
constraints for the SIS halo model in
Fig.~\ref{fig:sim-tsis-1gal-162}, derived from the analysis of data
with and without the lensing signal, respectively.  One sees that the
parameter constraints derived in each case are quite similar,
suggesting that a SNLS-quality catalogue containing just 162 SNIa is
unlikely to constrain the halo properties. Indeed this is what one
might expect for a SNIa sample for which $\Delta\ln\mathcal{Z}=-0.6$,
indicating that the SIS halo model is marginally disfavoured relative
to the no-lensing model. 

In particular, we see that the 1D marginals
for $\sigma_\ast$ and $r_\ast$ both peak at zero (which corresponds to
no lensing signal), although the former does have a modest subsidiary
peak at the correct input value of $\sigma_\ast$ for the data
containing the lensing signal.  Another noteworthy feature is that,
whereas the value of $\eta$ is unconstrained, there is quite a strong
constraint restricting $\gamma$ to be negative. From (\ref{eq:trlaw}),
one sees that this corresponds to a rather curious truncation radius
scaling-law, for which galaxies with larger velocity dispersions have
smaller truncation radii. Moreover, provided $\eta$ is positive, this
also corresponds to more luminous (and presumably more massive)
galaxies having smaller truncation radii. This phenomenon was also
noted by J\"onsson et al. (2010b), but was erroneously interpreted as
being a constraint on the halo properties derived from the SNIa data,
which is clearly not the case, since it occurs even in the analysis of
simulations containing no lensing signal. Rather, from
(\ref{eq:sissigma}) and (\ref{eq:trlaw}), one sees that a negative
value of $\gamma$ allows for a smaller value of the truncation radius
$r_t$ to `offset' an increase in the velocity dispersion $\sigma$,
thereby reducing the lensing signal produced by the putative halo, as
required to be consistent with simulated data containing no lensing
signal.

In any Bayesian model selection analysis, it is instructive to
investigate the impact of our assumed priors, listed in
Table~\ref{tab:priors1}, on the value of $\Delta\ln\mathcal{Z}$. Since
we are concerned only with the relative evidence between the SIS halo
model and the model assuming no lensing, we need not consider the
priors on the nuisance parameters, since these are common to both
models.  For the halo parameters, the range of the uniform priors on
$\sigma_\ast$ and $r_\ast$ are reasonably uncontroversial and
motivated both by physical considerations and previous studies. For
the power-law exponents $\eta$ and $\gamma$, however, our prior
knowledge is far less certain and, as commented above, the constraints
on these parameters shown in Fig.~\ref{fig:sim-tsis-1gal-162} display
some unusual features. We therefore confine our attention to these
parameters.

\begin{figure}
\begin{center}
\includegraphics[width=0.99\columnwidth]{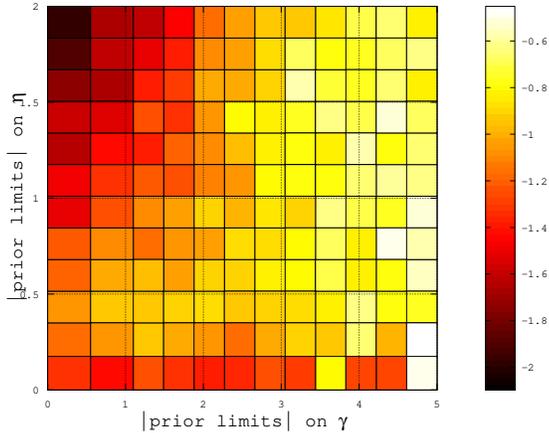}
\caption{Log-evidence difference $\Delta\ln\mathcal{Z}$ between the
  SIS halo model and the null for the `median' sample of 162 SNIa, as
  a function of the (symmetric positive and negative) limits on the
  uniform priors assumed for the parameters $\eta$ and $\gamma$.
\label{fig:sis162etagamma}}
\end{center}
\end{figure}

In Fig.~\ref{fig:sis162etagamma}, we plot $\Delta\ln\mathcal{Z}$ for
the `median' sample of 162 SNIa as a function of the (symmetric
positive and negative) limits on the uniform priors assumed for $\eta$
and $\gamma$.
In particular, one sees that as the prior limits on $\gamma$ are
increased, $\Delta\ln\mathcal{Z}$ also increases, indicating an
improvement in the quality of the fit to the data that outweighs the
Occam's razor penalty of widening the prior range. This behaviour,
although unusual, may be understood from the parameter constraints
plotted in Fig.~\ref{fig:sim-tsis-1gal-162}, which show a clear
preference for large negative values of $\gamma$, as discussed
above. Moreover, provided the prior on $\gamma$ is wider than about
$\pi(\gamma) = {\cal U}(-3,3)$, there is little dependence of
$\Delta\ln\mathcal{Z}$ on the width of the prior on $\eta$. For narrow
priors on $\gamma$, however, one sees that there is a broad peak in
$\Delta\ln\mathcal{Z}$ centred on priors for $\eta$ of around
$\pi(\eta)={\cal U}(-\tfrac{1}{2},\tfrac{1}{2})$, which is just
sufficient to encompass the posterior probability mass associated with
the true input value of $\eta=1/3$. Narrower priors on $\eta$ produce
smaller values of $\Delta\ln\mathcal{Z}$ since they do not contain
this probability mass, whereas wider priors on $\eta$ lead to smaller
smaller values of $\Delta\ln\mathcal{Z}$ since they incur an Occam's
razor penalty for increasing the extent of the prior without obtaining
a compensating improvement in the quality of the fit to the data.
Finally, we note that in all cases, $\Delta\ln\mathcal{Z}$ remains
negative, showing that the SIS halo model is disfavoured relative to
the no-lensing model.

\subsection{Results for 162 SNLS3-like supernovae}

To match the real SNLS3 data to be analysed in Section~\ref{sec:real},
we also constructed a simulation containing 162 SNIa, with positions
on the sky and redshifts fixed to those of the real 162 SNLS3
SNIa. The parameter constraints obtained in the analysis of data, with
and without a simulated lensing signal, are very similar to those
obtained for the `median' sample of 162 SNIa, shown in
Fig.~\ref{fig:sim-tsis-1gal-162}, so we do not plot them here. 

In this case, the Bayesian log-evidence for the SIS halo model,
relative to the null (no-lensing) model, is found to be
$\Delta\ln\mathcal{Z} = -1.3 \pm 0.2$ for the data set containing a
simulated lensing signal. Thus, according to the Jeffrey's scale, the
no-lensing model is `substantially' preferred (just) over the SIS halo
model.  This again suggests that the quantity and quality of the real
SNLS3 data to be analysed in Section~\ref{sec:real} are insufficient to
obtain a detection of the lensing signal assuming in the
simulation. Indeed, we note that value of $\Delta\ln\mathcal{Z}$
obtained lies near the low end of the range of those for the one
hundred random samples of 162 SNIa, as plotted in
Fig.~\ref{fig:sis162hist}. Nonetheless, when the limits on the uniform
priors for $\eta$ and $\gamma$ are allowed to vary, one obtains a
similar variation of $\Delta\ln\mathcal{Z}$ to that displayed in
Fig.~\ref{fig:sis162etagamma}; we therefore do not plot it here.

\subsection{Results for random samples of 500 supernovae}

To test that our analysis procedure is capable of detecting the
gravitational lensing signal and placing the correct constraints on
halo parameters in presence of more data, we also analyse simulated
random samples of 500 SNIa, which contain approximately three times
the number of SNIa as considered previously and is thus representative
of what could be achieved by the SLNS programme in a total of about 9
years of observation. This is clearly rather unrealistic in terms of
the required observing time and resources, but still provides a useful
insight into the quantity of SNIa data required to make a robust
detection of the lensing signal and constrain halo properties.

\begin{figure}
\begin{center}
\includegraphics[width=0.95\columnwidth]{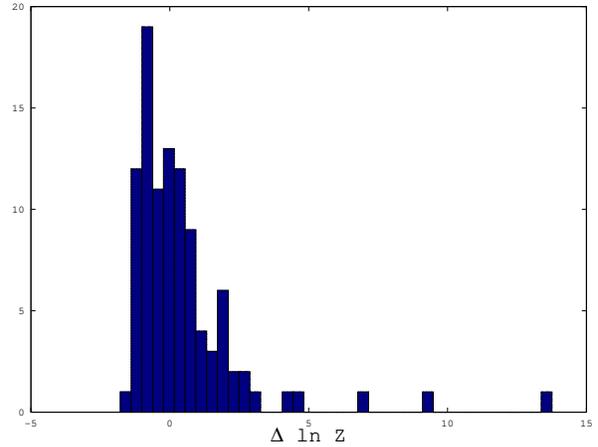}
\caption{Histogram of the log-evidence difference
  $\Delta\ln\mathcal{Z}$ between the SIS halo model and the null
  (no-lensing) model obtained from the analysis of one hundred
random samples of 500 SNIa.\label{fig:sis500hist}}
\end{center}
\end{figure}

In Fig.~\ref{fig:sis500hist}, we plot the histogram of the
log-evidence difference $\Delta\ln\mathcal{Z}$, for the SIS halo model
relative to the null (no-lensing) model, obtained from the analysis of
one hundred random samples of 500 SNIa.
Once again, the $\Delta\ln\mathcal{Z}$ values vary considerably
between samples, ranging from about $-2$ to $13$. The histogram has a
median $0.1$, mean $0.5$ and standard deviation $2.1$, showing that,
as expected, the distribution is shifted to larger values of
$\Delta\ln\mathcal{Z}$, as compared to the corresponding histogram in
Fig.~\ref{fig:sis162hist}, obtained from random samples of 162 SNIa.
Also, the high-end of the distribution contains a small number of
samples for which $\Delta\ln\mathcal{Z}$ is very large, indicating a
`decisive' preference for the SIS halo model over the no-lensing
model. Nonetheless, the lower end of the distribution extends as far
as that obtained for random samples of 162 SNIa, and contains samples
for which the SIS halo model is still `substantially' disfavoured.  We
again find that the key factor in determining the value of
$\Delta\ln\mathcal{Z}$ for a given sample in whether it contains some
of the most strongly lensed SNIa. Hence, one's ability to detect and
characterise the gravitational lensing signal from putative dark
matter haloes is less a matter of how many SNIa one observes, but
rather whether one's sample contains some highly magnified examples,
although, clearly, the chances of finding such SNIa increases with the
total number observed. Moreover, highly-magnified SNIa do not appear
to be concentrated at particularly high redshifts, so the importance
of observing high-redshift SNIa is not clear in this application.

\begin{figure}
\begin{center}
\includegraphics[width=0.95\columnwidth]{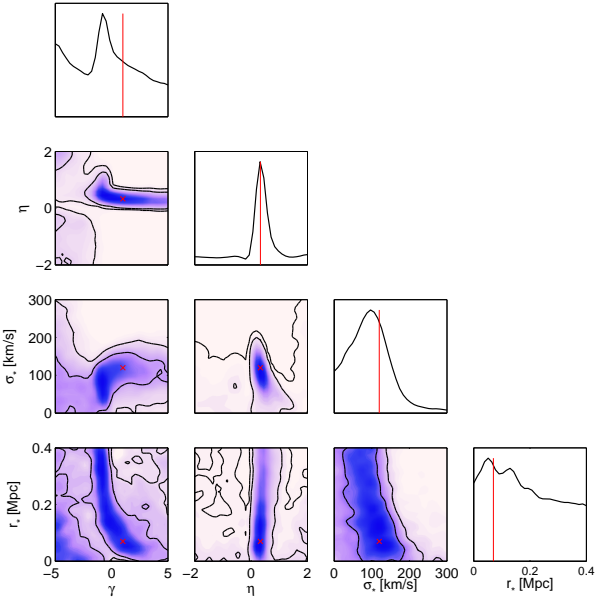}
\caption{1D and 2D marginalised posteriors distributions for the
  parameters $\bmath{h}=\{\gamma,\eta,\sigma_\ast,r_\ast\}$ of the
  truncated SIS halo model, derived from the analysis of 500 simulated
  SNIa data generated from a truncated SIS
  model (right). True parameter values are
  indicated by vertical lines and crosses in 1D and 2D plots,
  respectively. The SNIa sample used corresponds to that with the
  median value $\Delta\ln\mathcal{Z}=0.1$ from the histogram in
  Fig.~\ref{fig:sis500hist}. \label{fig:sim-tsis-1gal-500}}
\end{center}
\end{figure}

For the random sample that yields the median value of
$\Delta\ln\mathcal{Z}=0.1$, we plot the corresponding parameter
constraints for the SIS halo model in
Fig.~\ref{fig:sim-tsis-1gal-500}, derived from the analysis of data
with a lensing signal. For this random sample, although the SIS halo
model and the no-lensing model are essentially equally good
descriptions of the data, according to the Bayesian evidence, one sees
that most of the 1D and 2D marginalised posteriors have a well-defined
peak (away from zero) that contains the true parameter values,
indicating that reasonable constraints can be placed on the SIS halo
parameters. Nonetheless, some of the 2D marginals exhibit pronounced
degeneracies between the parameters, particular those involving the
parameter $r_\ast$, which is the least well-constrained
parameter. These features are consistent with the borderline value of
$\Delta\ln\mathcal{Z}$ obtained for this sample.

\begin{figure}
\begin{center}
\includegraphics[width=0.99\columnwidth]{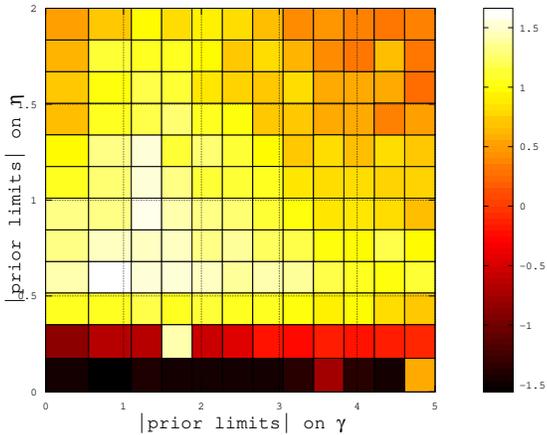}
\caption{Log-evidence difference $\Delta\ln\mathcal{Z}$ between the
  SIS halo model and the null for the `median' sample of 500 SNIa, as
  a function of the (symmetric positive and negative) limits on the
  uniform priors assumed for the parameters $\eta$ and $\gamma$.
\label{fig:sis500etagamma}}
\end{center}
\end{figure}

In Fig.~\ref{fig:sis500etagamma}, we plot $\Delta\ln\mathcal{Z}$ as a
function of the (symmetric positive and negative) limits on the
uniform priors assumed for $\eta$ and $\gamma$.
In this case, the resulting variation is broadly what might be
expected.  It exhibits a well-defined two-dimensional peak
corresponding roughly to the priors $\pi(\eta) = {\cal U}(-0.6,0.6)$
and $\pi(\gamma) = {\cal U}(-1.2,1.2)$, which are just wide enough to
encompass the posterior probability mass associated with the peak
centred on the true input values $\eta=1/3$ and $\gamma=1$. Indeed, for
such priors, the log-evidence difference is $\Delta\ln\mathcal{Z}
\approx 1.6$, indicating a `substantial' preference for the SIS halo
model. The evidence is lower both for narrower and wider priors on
each parameter, since such priors, respectively, either exclude this
probability mass or increase the prior volume without benefit.

\section{Application to real supernovae data}\label{sec:real}

We now apply our Bayesian analysis methodology to the real SNIa and
galaxies data sets described in Section~\ref{sec:data}. We also
investigate dividing the foreground galaxies in the SNLS catalogue
into passive and star-forming classes, each having SIS halo model
parameters that are allowed to be independent.

Assuming a single galaxy type ($\bmath{h}^{\rm p}=\bmath{h}^{\rm
  sf}$), the log-evidence for the SIS halo model, relative to the null
(no-lensing) model, is $\Delta\ln\mathcal{Z} = 0.2 \pm 0.2$,
indicating a very marginal preference for the former.  This is,
however, only at the level of the uncertainty in the evidence
calculation, and so this model and the no-lensing model are broadly
equally favoured.  When one splits the foreground galaxies into their
passive and star-forming spectral types ($\bmath{h}^{\rm
  p}\not=\bmath{h}^{\rm sf}$), one obtains the slightly larger value 
$\Delta\ln\mathcal{Z} = 0.5 \pm 0.2$, but this still corresponds 
only to a marginal preference for the SIS halo model. 

\begin{figure}
\begin{center}
\includegraphics[width=0.99\columnwidth]{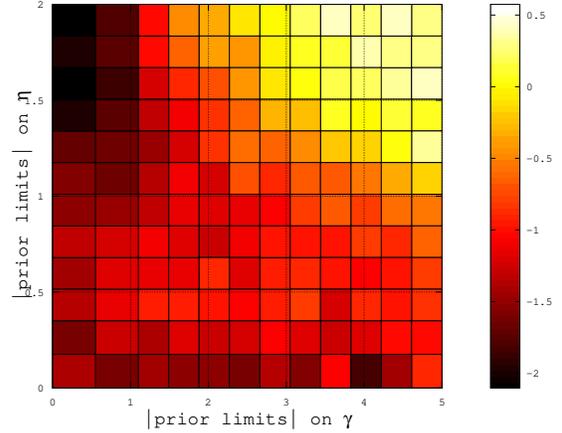}
\caption{Log-evidence difference $\Delta\ln\mathcal{Z}$ between the
  SIS halo model and the null for the real SNLS3 sample of 162 SNIa, as
  a function of the (symmetric positive and negative) limits on the
  uniform priors assumed for the parameters $\eta$ and $\gamma$.
\label{fig:sisrealetagamma}}
\end{center}
\end{figure}

In Fig.~\ref{fig:sisrealetagamma} we plot $\Delta\ln\mathcal{Z}$ as a
function of the (symmetric positive and negative) limits on the
uniform prior assumed for $\eta$ and $\gamma$, for the case in which
$\bmath{h}^{\rm p}=\bmath{h}^{\rm sf}$. We see that, for this real
data set, the resulting plot differs somewhat from that for the
`median' simulated sample containing 162 SNIa, given in
Fig.~\ref{fig:sis162etagamma}, but does have some features in common.
In particular, for priors on $\eta$ narrower than about $\pi(\eta) =
{\cal U}(-1,1)$, there is almost no dependence on the width of
$\pi(\eta)$ and only a very weak increase in $\Delta\ln\mathcal{Z}$ as
$\pi(\gamma)$ widens. By contrast, if the prior on $\eta$ is wider than
about $\pi(\eta) = {\cal U}(-1,1)$, there is a very strong dependence
of $\Delta\ln\mathcal{Z}$ on the width of both $\pi(\eta)$ and
$\pi(\gamma)$. The value of $\Delta\ln\mathcal{Z}$ falls rapidly with
increasing width of $\pi(\eta)$, whereas is grows rapidly as
$\pi(\gamma)$ widens. Indeed, it is only in the top right-hand corner
of the plot that $\Delta\ln\mathcal{Z} > 0$, indicating that the SIS
halo model is favoured over the no lensing model, albeit only
marginally.

We discuss the halo parameters constraints for the two models
($\bmath{h}^{\rm p}=\bmath{h}^{\rm sf}$ and $\bmath{h}^{\rm
  p}\not=\bmath{h}^{\rm sf}$), in the following subsections.

\subsection{Parameter constraints for a single galaxy type}

\begin{figure}
\begin{center}
\includegraphics[width=0.95\columnwidth]{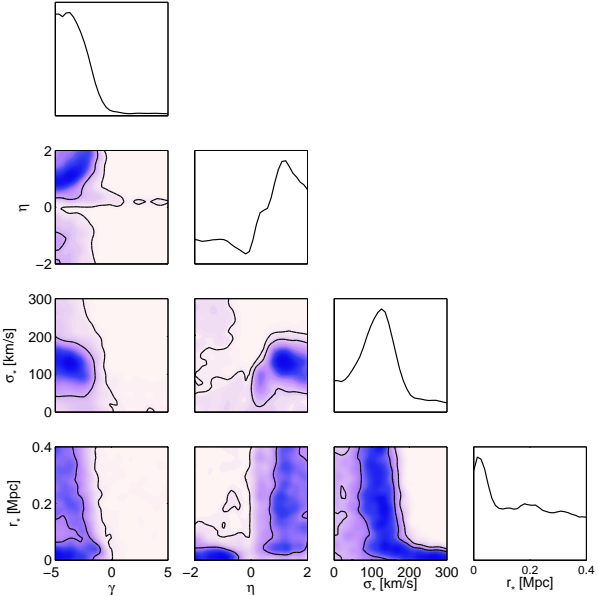}
\caption{1D and 2D marginalised posteriors distributions for the
  parameters $\bmath{h}=\{\gamma,\eta,\sigma_\ast,r_\ast\}$ of the
  truncated SIS halo model, derived from the analysis of real SNLS3 data.
  \label{fig:real-tsis-1gal}}
\end{center}
\end{figure}

The 1D and 2D marginalised parameter constraints obtained for the
truncated SIS halo model with one galaxy type are shown in
Fig.~\ref{fig:real-tsis-1gal}. 
We see that both the 1D and 2D constraints are somewhat tighter than
those plotted in Fig.~\ref{fig:sim-tsis-1gal-162} for the `median'
simulated sample of 162 SNIa, which were very similar for simulations
with and without a lensing signal. This may be indicative of a more
pronounced, albeit weak, lensing signal in the real data, although the
posterior distributions are still very broad. As might be expected,
the parameter constraints are significantly less well-defined than
those plotted in Fig.~\ref{fig:sim-tsis-1gal-500} for the `median'
simulated sample of 500 SNIa. It is worth noting, in particular, the
form of the 2D marginal in the $(\gamma,\eta)$-subspace, for which
this probability mass is concentrated in the top left-hand corner;
this corresponds to large positive values of $\eta$ and large negative
values of $\gamma$, which is consistent with the dependence of
$\Delta\ln\mathcal{Z}$ on the prior limits shown in
Fig.~\ref{fig:sisrealetagamma}.

Focussing on the 1D marginal posterior distribution of each halo
parameter, we first note the strong preference for negative values of
$\gamma$, as we observed in the analysis of simulations containing no
lensing signal. As explained in Section~\ref{sec:sim}, this is merely
a consequence of the lensing signal (if any) in the data being very
weak and does not constitute a meaningful constraint on the halo
properties.

Conversely, there is a relatively strong preference for positive
values of $\eta$, corresponding to a positive correlation between the
halo velocity dispersion and luminosity. Indeed, the marginal
distribution peaks at $\eta \approx 1$, which is considerably larger
than the canonical values of $\eta^{\rm p} = \tfrac{1}{4}$ and
$\eta^{\rm sf} = \tfrac{1}{3}$, corresponding to the Faber--Jackson
and Tully--Fisher relations, valid for early and late-type galaxies,
respectively. This constraint also differs noticeably from that shown
in the left-hand panel of Fig.~\ref{fig:sim-tsis-1gal-162}, obtained
from simulated data containing no lensing signal, which is suggestive
of a marginal lensing signal being present in the real data. 

The possibility of a faint lensing signal being present in the data is
also suggested by the 1D marginal distribution for the fiducial
velocity dispersion $\sigma_\ast$, which has a clear peak away from
zero. Indeed, the peak is centred on $\sigma_\ast \approx 120$ km
s$^{1}$, which was also the value assumed in the simulations analysed
in Section~\ref{sec:sim}. The fact that the parameter constraint
obtained from the real data again appears tighter than that obtained
from the simulations of 162 SNIa with no lensing signal, shown in the
left-hand panel of Fig.~\ref{fig:sim-tsis-1gal-162}, may be a result
of the real data preferring $\eta \approx 1$, as opposed the value
$\eta=\tfrac{1}{3}$ used in our simulations. 
From (\ref{eq:sissigma}) and
(\ref{eq:sislaw}), one sees that the lensing signal produced by a halo with a
given $\sigma_\ast$ is larger in the former case.

Finally, we note that the 1D marginal for the fiducial truncation
radius peaks very close to zero and yields no real constraint on
$r_\ast$, as we found in our analysis of the 162 simulated SNIa.

\subsection{Separating passive and star-forming galaxies}

So far we have assumed that the halo parameters for the truncated SIS
and NFW models, respectively, are the same for different galaxy
types. This may be an oversimplification and so it is of interest to
split the foreground galaxies into their passive and star-forming
spectral types, both of which are allowed to have independent halo
parameters.

The galaxies in our sample are classified as either passive or
star-forming depending on their sSFR (see
Section~\ref{sec:data:gal}). We thus allow all the halo parameters to be
different for each type of galaxy, so that the full halo parameter
space becomes $\bmath{h}=\{\gamma^{\rm p},\eta^{\rm p},\sigma^{\rm
  p}_\ast,r^{\rm p}_\ast,\gamma^{\rm sf},\eta^{\rm sf},\sigma^{\rm
  sf}_\ast,r^{\rm sf}_\ast\}$. The fraction of passive galaxies ranges
from 5 per cent in the CFHTLS fields D1 and D3 to 8 per cent in field
D4. Thus, the vast majority of foreground galaxies are star-forming.

\begin{figure*}
\begin{center}
\includegraphics[width=1.8\columnwidth]{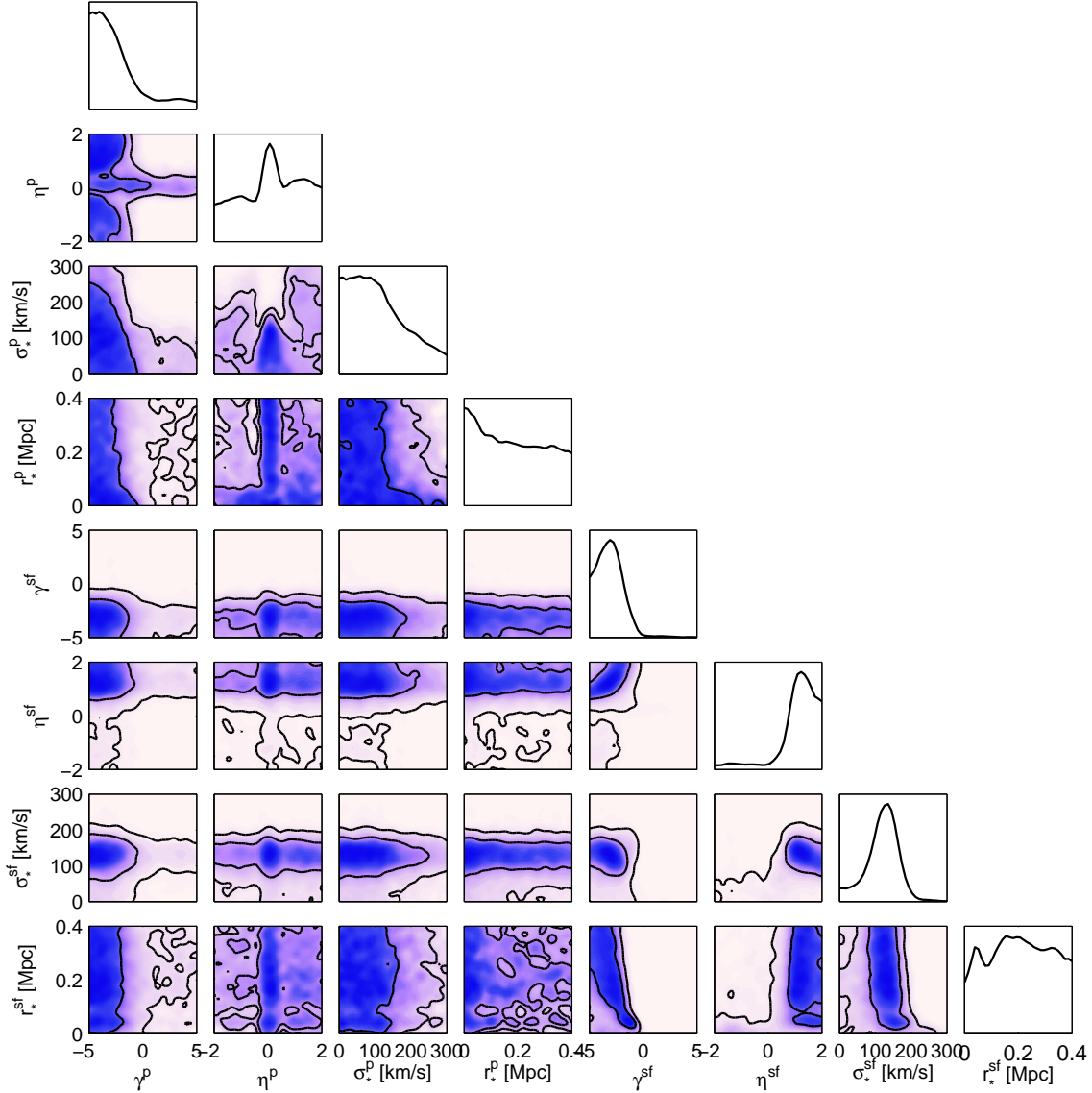}
\caption{1D and 2D marginalised posteriors distributions for the
  parameters $\bmath{h}=\{\gamma^{\rm p},\eta^{\rm p},\sigma^{\rm p}_\ast,r^{\rm
  p}_\ast,\gamma^{\rm sf},\eta^{\rm sf},\sigma^{\rm sf}_\ast,r^{\rm sf}_\ast\}$
   of the truncated SIS halo model for passive
  and star-forming galaxies, respectively, derived from the analysis
  of real SNLS3 data.
  \label{fig:real-tsis-2gal}}
\end{center}
\end{figure*}

The resulting 1D and 2D marginalised posteriors for the parameters are
shown in Fig.~\ref{fig:real-tsis-2gal}.  One sees that the constraints
on the parameters for passive and star-forming galaxies are very
different. As one might expect, given the relative percentages of
passive and star-forming galaxies in the catalogue, the constraints on
the halo parameters of the passive galaxies are very much weaker than
those for star-forming galaxies. Indeed, for the passive galaxies, the
marginal distributions closely resemble those shown in
Fig.~\ref{fig:sim-tsis-1gal-162}, obtained from simulations containing
162 SNIa with no lensing signal. Conversely, for star-forming
galaxies, the constraints are slightly tighter than those shown in
Fig.~\ref{fig:real-tsis-1gal}, obtained from the real data assuming
just one galaxy type, but resemble them in their main features. In
particular, we again see a strong preference for negative values of
$\gamma^{\rm sf}$, but now the 1D marginal distribution has a modest
peak at $\gamma^{\rm sf} \approx -3$. We also recover a strong
constraint that $\eta^{\rm sf}$ is positive, with the 1D marginal
possessing a small peak at $\eta^{\rm sf} \approx 1$. The 1D marginal
for $\sigma^{\rm sf}_\ast$ peaks strongly away from zero, preferring a
value of $\sigma^{\rm sf}_\ast \approx 150$ km s$^{-1}$. There is,
however, no constraint on $r^{\rm sf}_\ast$, which is in keeping with
our previous findings.

Finally, it is worth noting that the 2D marginal in the
$(\sigma_\ast^{\rm p},\sigma_\ast^{\rm sf})$-space shown in
Fig.~\ref{fig:real-tsis-2gal} bears a passing resemblance to the
corresponding plot (figure 6) in \cite{jonssonSNLS}, but only if
$\sigma_\ast^{\rm p}$ and $\sigma_\ast^{\rm sf}$ are
interchanged. Given the relative percentages of passive and
star-forming galaxies in the catalogue, it seems sensible that one
should obtain a tighter constraint on $\sigma_\ast^{\rm sf}$ than
$\sigma_\ast^{\rm p}$, as we find in Fig.~\ref{fig:real-tsis-2gal}. By
contrast, figure 6 in \cite{jonssonSNLS} shows the opposite,
which suggests that they have erroneously swapped the galaxy types in
their analysis.

\subsection{Inclusion of additional low-redshift supernovae data}

We also investigate the inclusion of additional, low-redshift SNIa
data to our analysis as a potential means of enhancing the detection
of a halo lensing signal. The rationale here is first to analyse the
low-$z$ SNIa data alone, using precisely the same methodology as for
the SNLS3 data, but assuming no lensing from foreground galaxy
haloes. The resulting posterior distributions derived for (most of)
the nuisance parameters listed in Table~\ref{tab:priors1} may then
used as priors on these parameters in the subsequent analysis of the
(high-$z$) SNLS3 data, thereby replacing the very conservative priors
assumed in Table~\ref{tab:priors1}, which may be `diluting' the
lensing signal.

As our low-$z$ SNIa dataset, we use an updated version of the
catalogue compiled by \citet{2010MNRAS.406..782S}. This consists of a
sample from the compilation of  \citet{conley11}, which itself
includes SNIa from a variety of sources (primarily \cite{1996AJ....112.2391H}; \cite{1999AJ....117..707R}; \cite{2006AJ....131..527J}; \cite{2009ApJ...700.1097H};
\cite{2010AJ....139..519C}). \citet{2010MNRAS.406..782S}  apply bulk-flow peculiar
velocity corrections to the SN magnitudes and redshifts, placing the
redshifts in the CMB-frame ($z_{\rm cmb}$) following \citet{2007ApJ...661L.123N}, but with updated models (\cite{conley11}). Only SN Ia in
the smooth Hubble flow, defined as $z_{\rm cmb} > 0.01$ are used, and
the same light curve quality cuts as for the SNLS3 sample are
employed. There are 123 low-redshift SN Ia in total, with redshifts in
the range $[0.01, 0.08]$.

Analysis of these data yields approximately Gaussian posteriors
(truncated to positive values) on the nuisance parameters, given by:
$\alpha \sim {\cal N}(1.30,0.11^2)$, $\beta \sim {\cal
  N}(2.81,0.18^2)$, $R_x \sim {\cal N}(0.119,0.008^2)$ and $R_c \sim
{\cal N}(0.074,0.05^2)$. These distributions are clearly much tighter
than the original assumed priors given in Table~\ref{tab:priors1};
note that we use the original prior on the nuisance parameter
$\sigma_{\rm int}$, as this quantity is likely to differ between
low-$z$ and high-$z$ SNIa, since observational uncertainties such as
the estimation of photometric errors can lead to variations of
$\sigma_{\rm int}$ on a sample-by-sample basis \citep{Paper1}. We find
that using the posteriors derived from the low-$z$ SNIa as priors in
the subsequent analysis of the high-$z$ SNLS3 data yields parameter
constraints that are almost identical to those presented in
Figs~\ref{fig:real-tsis-1gal}--\ref{fig:real-tsis-2gal}. Moreover, the
log-evidences relative to the null (no-lensing) model are found to be
very similar to those obtained previously. This finding makes sense,
since the new, tighter priors on the nuisance parameters are used for
both the halo and null (no-lensing) models, and they are, in fact,
consistent with the posteriors derived on these parameters from the
SNLS3 data alone. Thus, in summary, our results are unchanged by the
inclusion of low-$z$ SNIa data.

\section{Conclusions}\label{sec:con}

We have presented a Bayesian statistical methodology for constraining
the properties of dark matter haloes of foreground galaxies that
intersect the lines-of-sight towards SNIa. The method builds upon the
Bayesian hierarchical model presented by \citet{march11} for
improving constraints on cosmological parameters from SNIa
observations. Compared with the usual $\chi^2$-method, which suffers
from shortcomings in terms of its statistical foundations and
robustness, \citet{march11} demonstrate that the Bayesian method
delivers tighter statistical constraints, reduces statistical bias and
produces confidence intervals with better statistical coverage. 

We use this methodology to obtain an effective likelihood
function giving the probability of obtaining the observed SNIa data
(i.e. the parameter values obtained in \salt{} lightcurve fits) as a
function of the parameters of the dark matter halo models assumed for
the galaxies along the lines-of-sight to the SNIa. Following the
imposition of suitable priors on these parameters (together with some
nuisance parameters), we explore the full posterior distribution in
all the parameters simultaneously using the nested sampling algortihm
{\sc MultiNest}, which also calculates the Bayesian evidence for use
in model comparison.

We first apply our method to simulated SNIa datasets generated using
162 high-redshift ($0.1 \la z \la 1$) SNIa from the 3-year data
release of the Supernova Legacy Survey (SNLS3) as a template and
assuming a truncated singular isothermal sphere (SIS) model for the
dark matter halo density profile of foreground galaxies in the deep
Canada-France-Hawaii Telescope Legacy Survey (CFHTLS) fields. These
simulations were generated using the SNANA package, assuming realistic
values for the halo parameters and observational data quality.
Assuming conservative priors on the parameters, we demonstrate that
there is a wide variation in the significance at which one may detect
a gravitational lensing signal, depending on whether the sample
contains some SNIa that are strongly magnified. Indeed, the
log-evidence $\Delta\ln\mathcal{Z}$, relative a model assuming no
lensing, ranges from from about $-1.5$ to $4.5$, with a median value
of $-0.6$. For this median catalogue, the parameter constraints are
very broad and resemble those obtained from analysing a simulation
containing no lensing signal.

Analysing simulated catalogues containing 500 SNIa, we again find a
wide variation in the significance to which one may detect the lensing
signal, depending on the number of highly-magnified SNIa contained in
the sample. Nonetheless, as might be expected in this case, the
distribution of $\Delta\ln\mathcal{Z}$ is shifted to larger values,
ranging from about $-2$ to $13$, with a median value of $0.1$. For
this median sample, our method produces posterior distributions for
the parameters that have well-defined peaks, which contain the true
input values used in the simulations. This demonstrates that our method
can indeed detect the lensing signal and estimate the halo parameter
values correctly, provided one analyses a sufficient number of SNIa.

In the analysis of real SNLS3 data (consisting of 162 SNIa) we find,
contrary to previous studies, only a very marginal detection of a
lensing signal in the case of the truncated SIS halo model. Assuming
conservative priors on the halo parameters, the model is preferred by
just 0.2 log-evidence units relative to the no lensing model. Indeed,
since this difference is similar to the uncertainty in the evaluation
of the evidence, one may consider the no lensing model to be equally
favoured by the data. We also show that assuming narrower priors
centred on zero for the exponents $\eta$ and $\gamma$ of the assumed
power-law relations between velocity dispersion, luminosity and
truncation radius leads to {\em smaller} log-evidence values.
Nonetheless, the parameter constraints for the truncated SIS halo
model do appear somewhat tighter than those obtained for simulations
of 162 SNIa without the inclusion of a lensing signal, which is again
suggestive of a borderline detection of a lensing signal in the real
SNLS3 data. One finds that the SNLS3 data strongly prefer negative
values of $\gamma$, which corresponds to luminous galaxies having
smaller truncation radii than less luminous ones, but this is simply a
manifestation of the lensing signal being very weak and does not
constitute a meaningful constraint on the halo properties. There is
also a preference for positive values of $\eta$, corresponding to a
positive correlation between the halo velocity dispersion and
luminosity, with the 1D marginal for this parameter peaking at $\eta
\approx 1$, which is somewhat larger than the canonical values of
$\eta^{\rm p} = \tfrac{1}{4}$ and $\eta^{\rm sf} = \tfrac{1}{3}$ valid
for early and late-type galaxies, respectively, and leads to a
stronger lensing signal for a halo with a given luminosity. The 1D
marginal distribution for the fiducial velocity dispersion
$\sigma_\ast$ has a clear peak away from zero, which is centred on
$\sigma_\ast \approx 120$ km s$^{-1}$, but the marginal for the
fiducial truncation radius peaks very close to zero and yields no real
constraint on $r_\ast$. 

Finally, we investigate the possibility that the halo parameters may
be different for passive and star-formation galaxies,
respectively. Focussing on the truncated SIS model, we find that
allowing all the halo parameters for the two galaxy types to be
completely independent increases the evidence for the model slightly,
by 0.3 log-evidence units. Since only around 5 per cent of the
galaxies in our catalogue are passive, we find that one cannot place
any meaningful constraints on their halo properties, but that the
constraints on the halo parameters of the star-forming galaxies are
similar to  those obtained if one assumes just one galaxy type.

Our results contradict to some extent the previously reported
high-significance detections of gravitational lensing of SNIa
\citep{jacob07,kronborg10,jonssonGOODS,jonssonSNLS}, where the last
study uses essentially the same SNLS3 data as those analysed here. We
have also verified that our findings are unchanged by the inclusion of
additional low-$z$ SNIa data. The major difference between these
earlier analyses and the study presented here is the statistical
methodology employed.  As demonstrated by \citet{march11}, the usual
$\chi^2$-method used in previous analyses has a number of shortcomings
in terms of its statistical foundations and robustness, including not
allowing for rigorous model checking and not providing a reliable
framework for the evaluation of systematic uncertainties. This should
be contrasted with the statistically-principled and rigorous Bayesian
hierarchical model used here.  Moreover, previous studies employed
using simple gridding methods to explore the parameter space of the
dark matter halo models, which allow only a small subset of the
parameters to be varied simultaneously. In particular,
\cite{kronborg10} fixes all the halo parameters to `reasonable' values
to derive a statistical significance for the presence of a lensing
signal.

It should be mentioned that the analysis presented here does make the
significant simplifying assumption that all dark matter haloes of the
foreground galaxies have the same values for their free parameters (at
least within their spectral type). This is not too restrictive an
assumption, however, for free parameters corresponding to fiducial
values appearing in scaling laws, but it is certainly a
oversimplification.  This may weaken the constraints in the analysis of
real data, as compared with the analysis of simulated data generated
assuming all the haloes to have the same parameter values. We will
investigate this possibility in a future work

In closing, we note that the analysis code used in this paper is
quite general in nature and could easily be applied to other
combinations of SNIa and foreground galaxy catalogues. Anyone wishing to
use the code in collaboration should contact the authors. 

\section*{Acknowledgements}\label{sec:ackn}
NVK thanks Jakob J\"onsson for his invaluable help in the initial
stages of this work, Mark Sullivan for providing the foreground galaxy
catalogues and for many useful discussions, Ariel Goobar for reading
through an early draft of the paper and providing insightful comments,
and Joakim Edsj\"o for his support and encouragement throughout.  NVK
also acknowledges support from the Swedish Research Council (contract
No. 621-2010-3301). MCM thanks Rick Kessler and John Marriner for
assistance with simulating and analyzing SNANA data. FF is supported
by a Research Fellowship from Trinity Hall, Cambridge.  This work was
performed on COSMOS VIII, an SGI Altix UV1000 supercomputer, funded by
SGI/Intel, HEFCE and PPARC.

\bibliographystyle{mn2e}
\bibliography{references}

\label{lastpage}
\end{document}